\documentclass[a4paper,11pt,final]{article}

\usepackage{graphicx,amsmath,amssymb,algorithm,algorithmic,textcomp,tikz-cd,float} 
\usepackage{epsfig}

\begin{document}

\title{\ \\ \LARGE\bf Relationships between dilemma strength and fixation properties in coevolutionary games}

\author{Hendrik Richter \\
HTWK Leipzig University of Applied Sciences \\ Faculty of
Electrical Engineering and Information Technology\\
        Postfach 301166, D--04251 Leipzig, Germany. \\ Email: 
hendrik.richter@htwk-leipzig.de }

\maketitle

\begin{abstract}
Whether or not cooperation is favored over defection in evolutionary games can be assigned by structure coefficients for any arrangement of cooperators and defectors  
on any network modeled as a  regular graph. We study how these structure coefficients relate to 
a scaling of dilemma strength in social dilemma games. It is shown that some graphs  permit certain arrangements of cooperators and defectors to possess particularly large structure coefficients. Moreover, these large coefficients imply particularly large sections of a bounded parameter plane spanned by scaling  gamble--intending and risk--averting dilemma strength.

\end{abstract}

\section{Introduction}
A fundamental issue in understanding evolutionary dynamics of biological systems is the interplay between competition and  cooperation. Evolutionary dynamics requires Darwinian selection in which biological entities compete in terms of survival and reproduction. But while we certainly observe competition in living entities,  juxtaposed and intertwined with it, we also notice very frequently unselfish,  altruistic and cooperative behavior. The paradox of competition and cooperation next to each other can be  resolved by presuming that there must be situations where  in  evolutionary terms cooperation is more advantageous than competition. Mathematical models for  discussing these questions are provided by evolutionary game theory, which gives a theoretical framework and a bio--inspired computational paradigm~\cite{broom13,nowak06}. 

Studying the emergence of cooperation crystallizes into a well--defined form by considering so--called social dilemma games. In an evolutionary type of these games a population of players interacts in a predefined manner among themselves by each selecting one of two strategies, cooperate or defect, and receiving a payoff according to these selections. By converting the payoff into accumulable fitness, repeating the interaction and allowing players to change strategies depending on the accumulated fitness, the long--term effect of strategy selection becomes visible~\cite{allen17,chen13,hinder15,taylor07}. A frequently studied question of considerable biological relevance is whether or not one strategy is favored over another depending on the values of the payoff matrix \emph{and} on the structure of the interaction network specifying who--plays--whom~\cite{chen16,hinder15,rich18a,rich18b}.
Recently, and independent from each other, two proposals have been made to formalize strategy selection and payoff allocation, on the one hand, and strategy distribution over  interaction networks, on the other hand. Wang et al.~\cite{wang15} introduced an approach of universal scaling for payoff matrices that facilitates to study a continuum of social dilemmas including (but not restricted to) well--known examples such as the Prisoner’s dilemma (PD), the stag hunt (SH) and the snow drift (SD) game. Chen et al.~\cite{chen16} presented a scheme to define structure coefficients for any arrangement of cooperators and defectors on interaction networks modelled as regular graphs.
In this paper, these recent additions to evolutionary game theory are combined. There are simple algebraic relations between structure coefficients, the elements of any payoff matrix and whether or not cooperation is favored over defection~\cite{chen16,tarnita09}. Thus, we can study fixation properties across the universal scaling for payoff matrices and dilemma strength.   The paper is structured as follows. In Sec. 2 we briefly review coevolutionary games, particularly highlighting the scaling of dilemma strength~\cite{wang15} as well as configurations and structure coefficients~\cite{chen16,rich18a,rich18b}. Numerical results are presented for different interaction networks modeled as regular graphs. It is shown how different networks may entail different combinations of dilemma strength where cooperation is favored over defection.

\section{Game description}

\subsection{Coevolutionary games, payoff matrices and scaling for the dilemma strength }  \label{sec:desc}
We consider coevolutionary games of $N$ players $\mathcal{I}=\left(\mathcal{I}_1,\mathcal{I}_2,\ldots,\mathcal{I}_N\right)$.   The game has 2 strategies, cooperate ($C_i$) and defect ($D_i$) and each pairwise interaction between 2 players $\mathcal{I}_i$ and $\mathcal{I}_j$, $i\neq j$, (which thus are mutual coplayers) yields a  payoff. A (possibly varying) interaction network describes which player interacts with whom, while the number of coplayers is the same for all players. We need 
 three entities for describing such a game:
(i) the network of interaction,  (ii) the configuration describing the strategy of each player, and (iii) the payoff matrix~\cite{perc10,rich16,rich17}.




\begin{figure}[t]
\centering
\includegraphics[trim = 17mm 120mm 138mm 120mm,clip, width=4cm, height=4.2cm]{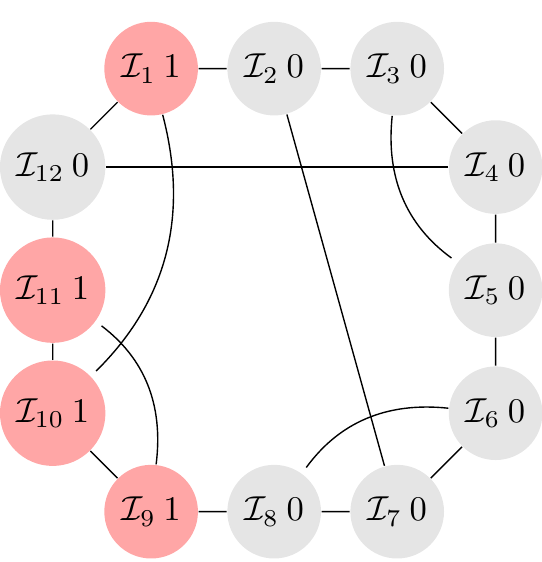} 
\includegraphics[trim = 17mm 120mm 138mm 120mm,clip, width=4cm, height=4.2cm]{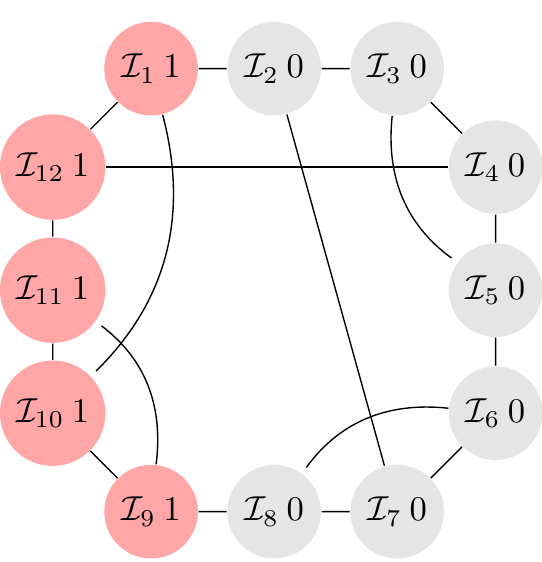} 
\includegraphics[trim = 17mm 120mm 138mm 120mm,clip, width=4cm, height=4.2cm]{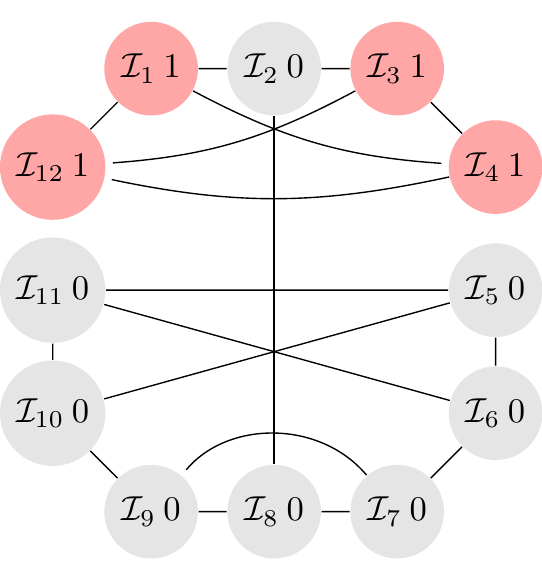} 
\includegraphics[trim = 17mm 120mm 138mm 120mm,clip, width=4cm, height=4.2cm]{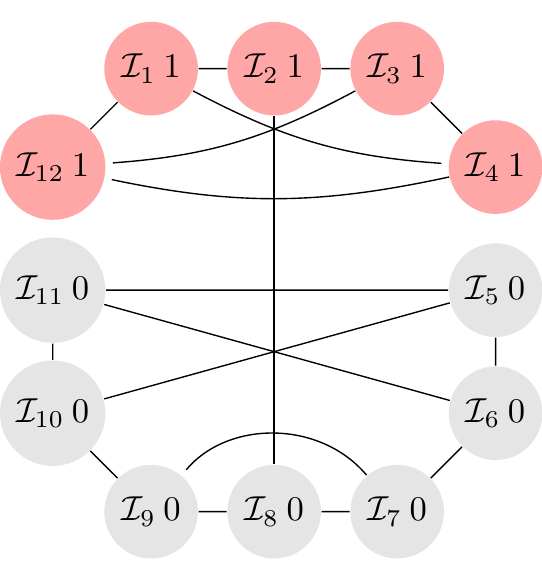}

 (a) $\sigma_{max}=1.7059$  \hspace{1cm} (b)  $\sigma_{max}=1.7568$ \hspace{1cm}  (c)  $\sigma_{max}=1.8485$ \hspace{1cm}  (d)  $\sigma_{max}=1.9159$

\includegraphics[trim = 17mm 120mm 138mm 120mm,clip, width=4cm, height=4.2cm]{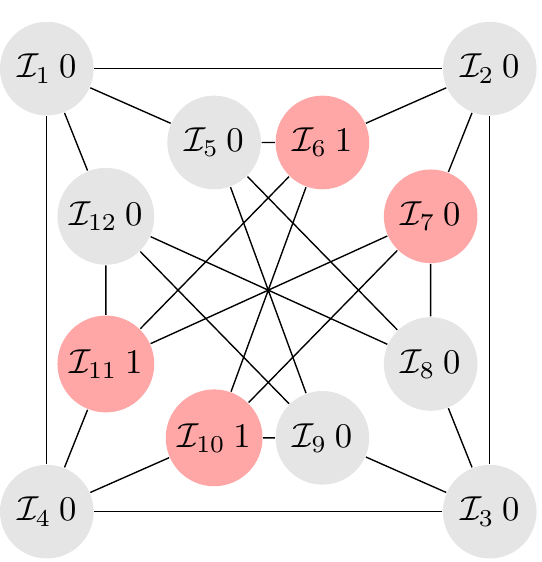} 
\includegraphics[trim = 17mm 120mm 138mm 120mm,clip, width=4cm, height=4.2cm]{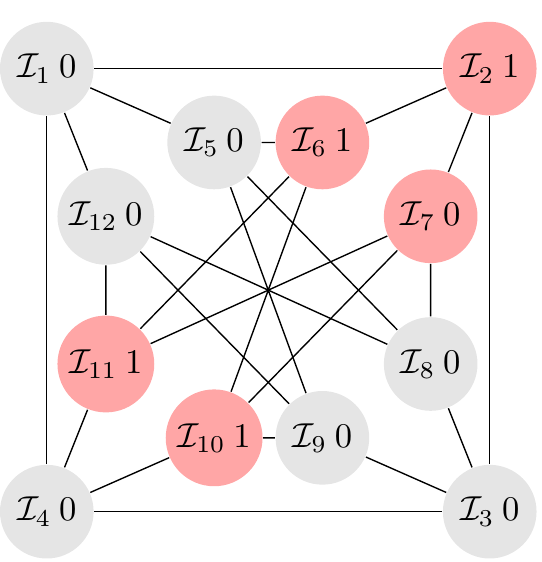} 
\includegraphics[trim = 17mm 120mm 138mm 120mm,clip, width=4cm, height=4.2cm]{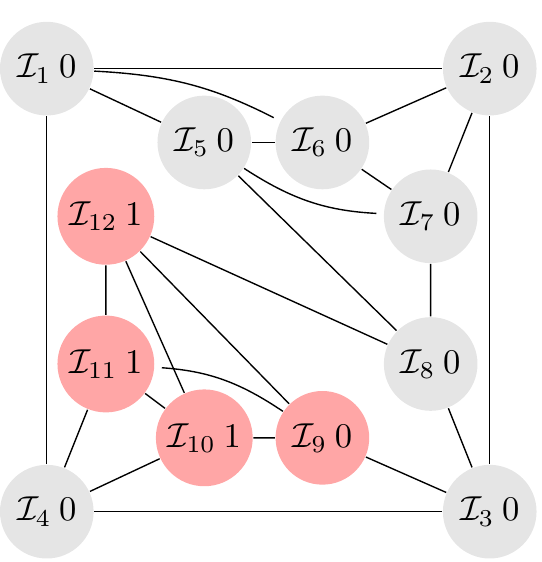} 
\includegraphics[trim = 17mm 120mm 138mm 120mm,clip, width=4cm, height=4.2cm]{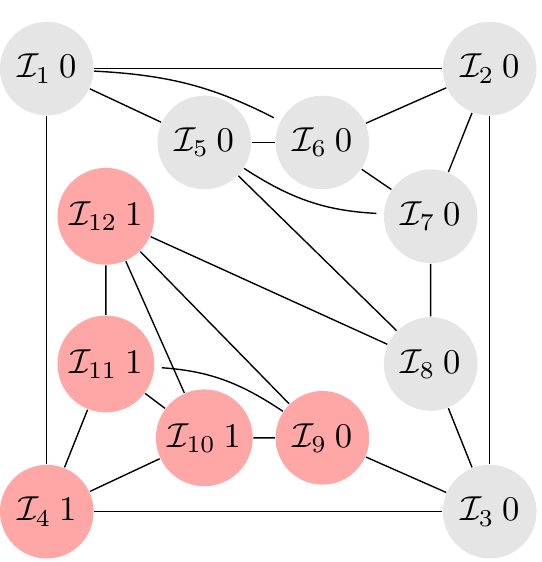}

 (e) $\sigma_{max}=1.2857$  \hspace{1cm} (f)  $\sigma_{max}=1.2957$ \hspace{1cm}  (g)  $\sigma_{max}=1.4433$ \hspace{1cm}  (h)  $\sigma_{max}=1.4727$
 
\caption{\small{Examples of interaction networks: (a)--(d) cubic ($3$--regular) and (e)--(h) quartic ($4$--regular) graphs. Two graph are named: (a),(b) the Frucht graph, (e),(f)  the Chvatal graph. The other two graphs have particularly large maximal structure coefficients $\sigma(\pi)=\sigma_{max}$ for some configurations $\pi$. A red vertex is a cooperator ($\pi_i=1$), while a grey vertex is a defector ($\pi_i=0$). The values of $\sigma_{max}$ are shown for: (a),(c),(e),(f) $c(\pi)=4$ cooperators; (b),(d),(f),(h) $c(\pi)=5$ cooperators. }}
\label{fig:graph}
\end{figure}

The interaction network is given by an undirected graph $\mathcal{G}=(V,E)$ and specifies who--plays--whom. According to evolutionary graph theory~\cite{allen14,sha12,szabo07}, the set of vertices $V$  equals the set of players $\mathcal{I}$ and the set of edges $E$ shows which  players  are mutual coplayers, see Fig. \ref{fig:graph} for examples.  As each players has $k$ coplayers, the interaction network is a $k$--regular graph.  The configuration $\pi=(\pi_1\pi_2\ldots\pi_N)$ specifies the strategy $\pi_i \in \{C_i,D_i\}$ of each player $\mathcal{I}_i$, ($i=1,2,\ldots,N$). With 2 strategies ($C_i$ and $D_i$) there are $2^N$ configurations.  These configurations enumerate all possible arrangements of cooperators and defectors among the players. Additionally, the configurations  describe  any outcome of a player changing its strategy in a strategy updating process, for instance death--birth (DB) or birth--death (BD) updating~\cite{allen14,sha12,zuk13}. 
It is convenient to binary code the strategies $\{C_i,D_i\} \rightarrow \{1,0\}$, thus having a binary string to specify the strategies of all players~\cite{chen16,rich17}. As an example, see Fig. \ref{fig:graph}(a) with the configuration  $\pi=(1000\:0000\:1110)$ showing players $\mathcal{I}_1$,$\mathcal{I}_9$, $\mathcal{I}_{10}$ and $\mathcal{I}_{11}$ cooperating, while the remaining 8 players defect.

\begin{figure}[t]

\includegraphics[trim = 3mm 0mm 0mm 0mm,clip, width=8.5cm, height=5.9cm]{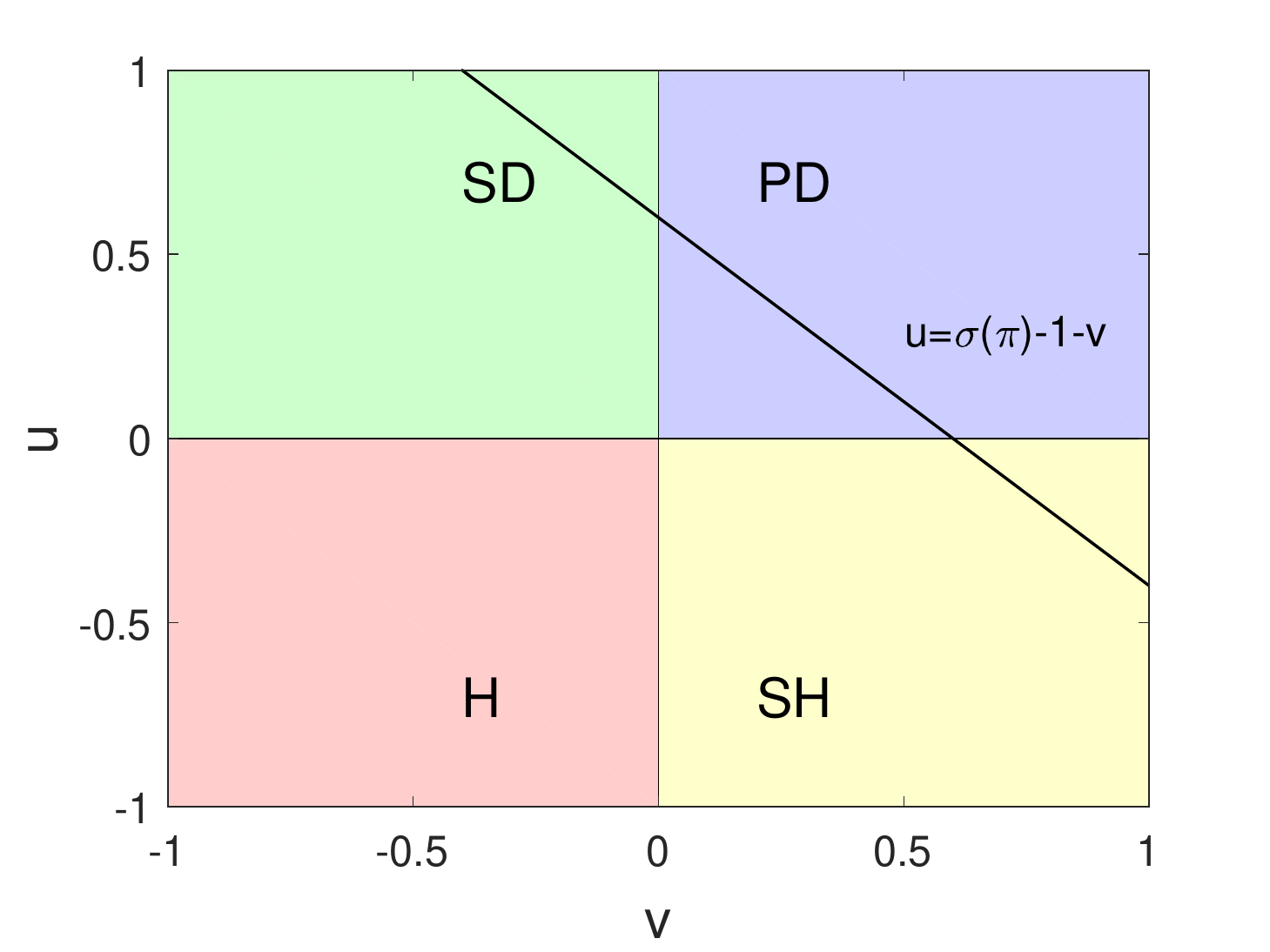} 
\includegraphics[trim = 3mm 0mm 0mm 0mm,clip, width=8.5cm, height=5.9cm]{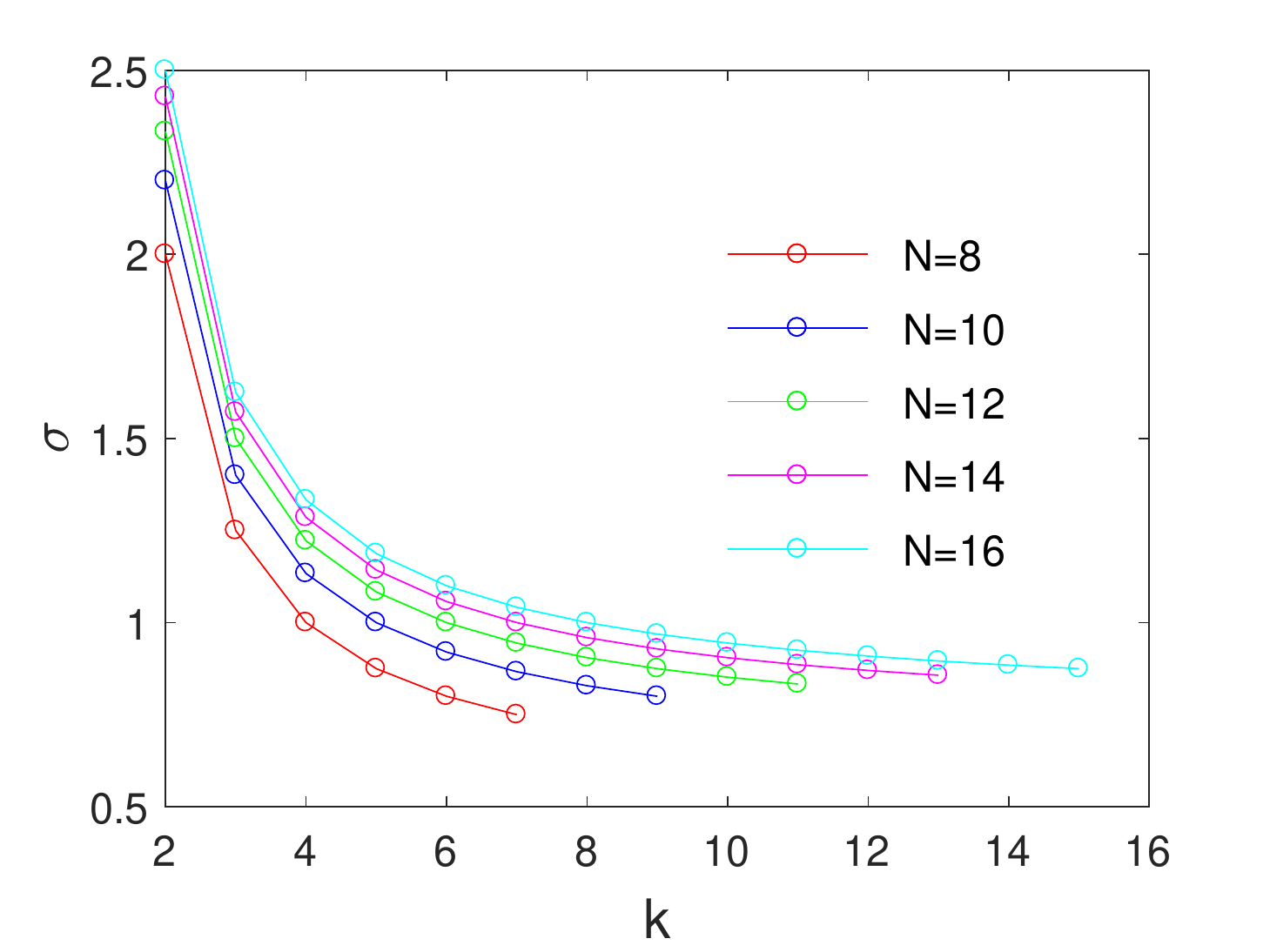} 

\hspace{1cm} (a)  \hspace{10cm} (b) 

\caption{(a) Schematic description of social dilemmas in a $uv$--parameter plane, as defined by the rescaled payoff matrix (\ref{eq:payoff1}). The plane can be divided into four sections (identified by different colors) which correspond to  prisoner's dilemma (PD),  snowdrift (SD), stag hunt (SH),  and harmony (H) games. Condition (\ref{eq:cond1}) implies straight lines $u=\sigma(\pi)-1-v$ from north--west to south--east in the $uv$--plane for which cooperation is favored over defection below these lines.   (b) The  structure coefficient of a single cooperator (\ref{eq:sig}) for different number of players $N$ over the number of coplayers $k$. The structure coefficient $\sigma$ falls rapidly for $k$ increasing, which means an increasing number of coplayers reduces the sections in $uv$--plane where  cooperation is favored over defection, see Fig. \ref{fig:uv_plane}(a).}
\label{fig:uv_plane}
\end{figure}

The $2 \times 2$ payoff matrix is 
\begin{equation} 
\bordermatrix{~ & C_j & D_j \cr
                  C_i & R & S \cr
                  D_i & T & P \cr} \label{eq:payoff}
\end{equation}
where $T$ is temptation to defect, $R$ is reward for mutual cooperation, $P$ is punishment for mutual defection, and $S$ is sucker payoff for cooperating with a defector. According to the values and order of these 4 elements of the payoff matrix (\ref{eq:payoff}), we obtain different social dilemma games.
Several suggestions have been made to rescale the payoff matrix (\ref{eq:payoff}) by freezing or linearly coupling its elements, which may reduce the $4$--dimensional parameter space to a $2$--dimensional plane~\cite{tani09,wang15,zuk13}, while preserving frequently--studied social dilemmas such as prisoner's dilemma (PD),   snowdrift (SD), stag--hunt (SH), or harmony (H). Following Wang et al.~\cite{wang15}, we consider two scaling parameters for the dilemma strength $u$ and $v$ to obtain a rescaled payoff matrix
\begin{equation} 
\bordermatrix{~ & C_j & D_j \cr
                  C_i & R & P-(R-P)v \cr
                  D_i & R+(R-P)u & P \cr} \label{eq:payoff1}
\end{equation}
with $u=\frac{T-R}{R-P}$ and $v=\frac{P-S}{R-P}$. We may interpret $u$ as gamble--intending dilemma strength and $v$ as risk--averting dilemma strength. The rescaling (\ref{eq:payoff1}) requires $R>P$, while $T-R$ and $P-S$ may change sign to have different orders of $(T,R,P,S)$, and thus different social dilemmas. Apparently, matrix (\ref{eq:payoff1}) reduces to matrix (\ref{eq:payoff}) by inserting $u$ and $v$.  However, by varying $u$ and $v$ for $-1 \leq u \leq 1$ and $-1 \leq v \leq 1$, we may traverse a bounded two--dimensional $uv$--parameter plane encompassing all the social dilemmas given above, but also some intermediate forms, see Fig. \ref{fig:uv_plane}(a). We obtain   SD games for  $0 \leq u \leq 1$  and  $-1 \leq v \leq 0$, PD games for $0 \leq u \leq 1$  and  $0 \leq v \leq 1$, and so on.  Thus, a rescaling by matrix (\ref{eq:payoff1})  significantly eases analyzing the games across social dilemmas. A square in the $uv$--plane generalizes the payoff matrix (\ref{eq:payoff}) and produces a multitude of dilemmas that are significant and interesting in evolutionary game theory. Moreover, Wang et al.~\cite{wang15} have shown that by the rescaling (\ref{eq:payoff1}) fixation properties of the games over the $uv$--plane are fairly robust with respect to the choice of $R$ and $P$.

\subsection{Fixation properties, configurations and structure coefficients}

Recently, Chen et al.~\cite{chen16} have shown that  for  $2 \times 2$ games with $N$ players, payoff matrix (\ref{eq:payoff}), any configuration $\pi$ of cooperators and defectors and for any interaction network modeled by a simple, connected, $k$--regular graph,
in the case of weak selection strategy $C_i$ is favored over $D_i$ if \begin{equation} \sigma(\pi)> \frac{T-S}{R-P}. \label{eq:cond} \end{equation} 
The quantity $\sigma(\pi)$ is called the structure coefficient of the configuration $\pi$ implying that it may have different values for different arrangements of cooperators and defectors described by $\pi$. It generalizes the structure coefficient \begin{equation} \sigma=\frac{(k+1)N-4k}{(k-1)N} \label{eq:sig} \end{equation} applying to a single cooperator~\cite{chen16,tarnita09}, see Fig. \ref{fig:uv_plane}(b) with curves of $\sigma$ for some $N$ and $k$. 
For the rescaled payoff matrix (\ref{eq:payoff1}) the condition (\ref{eq:cond}) simplifies to 
\begin{equation} \sigma(\pi)> 1+u+v.   \label{eq:cond1} \end{equation} Condition  (\ref{eq:cond1})  depends linearly on the scaling parameters for dilemma strength $-1 \leq u \leq 1$ and $-1 \leq v \leq 1$, but not on the parameter of the payoff matrix (\ref{eq:payoff}) and particularly not on $R$ and $P$. Obtaining such a simple algebraic expression as to whether or not 
cooperation is favored over defection over the whole $uv$--plane expressing all major social dilemmas is another advantage of the rescaling with payoff matrix (\ref{eq:payoff1}), as compared to other types of rescaling~\cite{tani09,zuk13}.
 
The structure coefficient $\sigma(\pi)$ can be calculated with time complexity  $\mathcal{O}(k^2N)$ for weak selection, the interaction graph $\mathcal{G}=(V,E)$ also describing the replacement structure,  and DB updating:  \begin{equation} 
\sigma(\pi)=\frac{N\left(1+1/k \right) \overline{\omega^1} \cdot \overline{\omega^0}-2\overline{\omega^{10}}-\overline{\omega^1 \omega^0} }{N\left(1-1/k \right) \overline{\omega^1} \cdot \overline{\omega^0}+\overline{\omega^1 \omega^0}}.  \label{eq:sigma}
\end{equation} The local frequencies  $\overline{\omega^1}$, $\overline{\omega^0}$, $\overline{\omega^{10}}$, $\overline{\omega^1 \omega^0}$ can be interpreted as follows~\cite{chen16,rich18a,rich18b}.  Suppose a random walk is carried out with the starting 
vertex chosen uniformly--at--random on a given interaction network. The local frequency $\overline{\omega^1}$ (or $\overline{\omega^0}=1-\overline{\omega^1}$) is the probability that for a configuration $\pi$ the player at the first step of the walk is a cooperator (or defector). The local frequency  $\overline{\omega^{10}}$ is the probability that for a walk with 2 steps the player at the first step is  a cooperator and at the second step it is a defector. The local frequency   $\overline{\omega^1 \omega^0}$ is the probability that for 2 random walks independent of each other the player at the first step on the first walk is a cooperator, but at the first step on the second walk is a defector.

\section{Numerical results and discussion}

\begin{figure}[htb]

\includegraphics[trim = 0mm 0mm 0mm 0mm,clip,width=8.5cm, height=5.9cm]{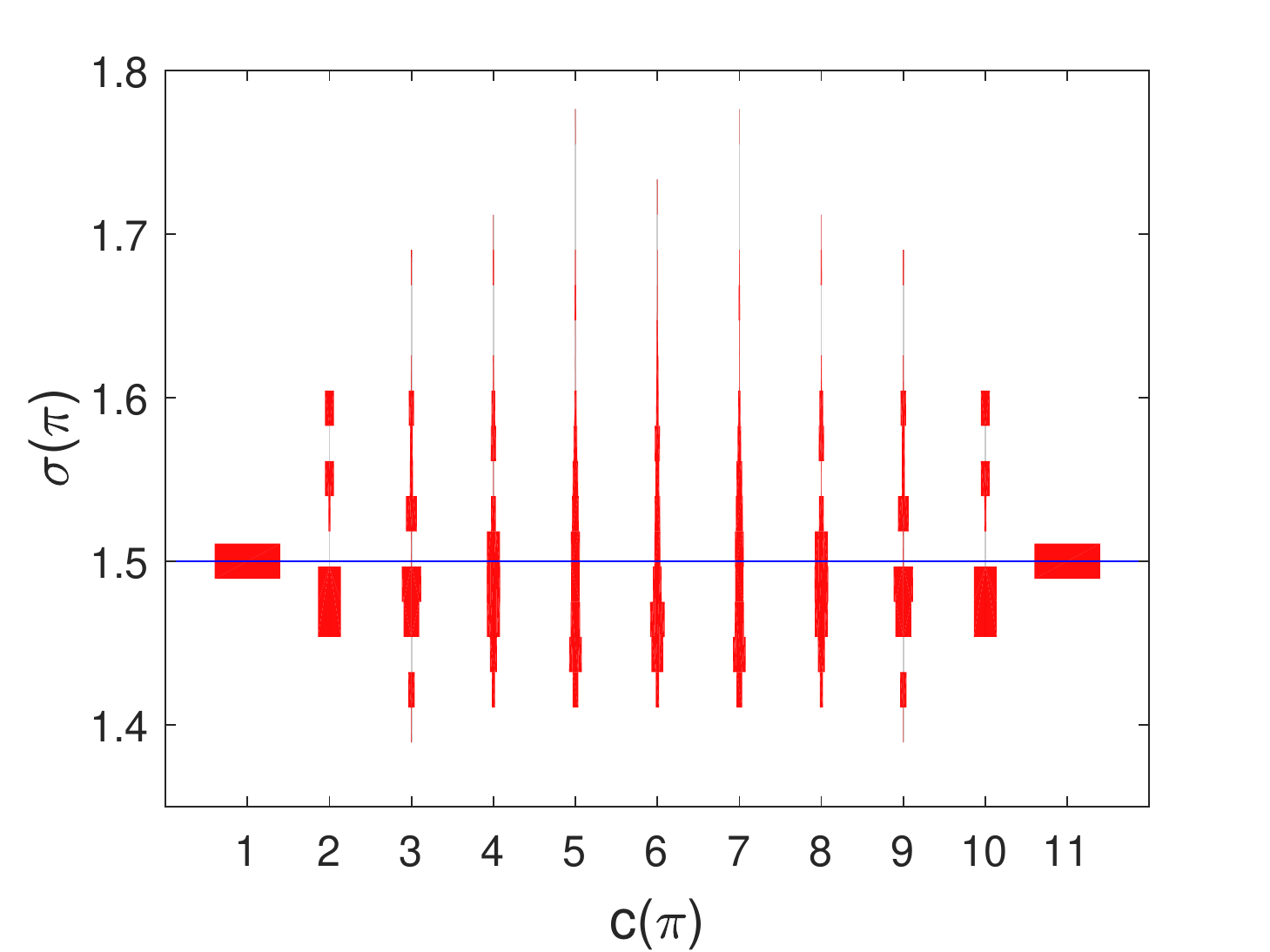}
\includegraphics[trim = 0mm 0mm 0mm 0mm,clip,width=8.5cm, height=5.9cm]{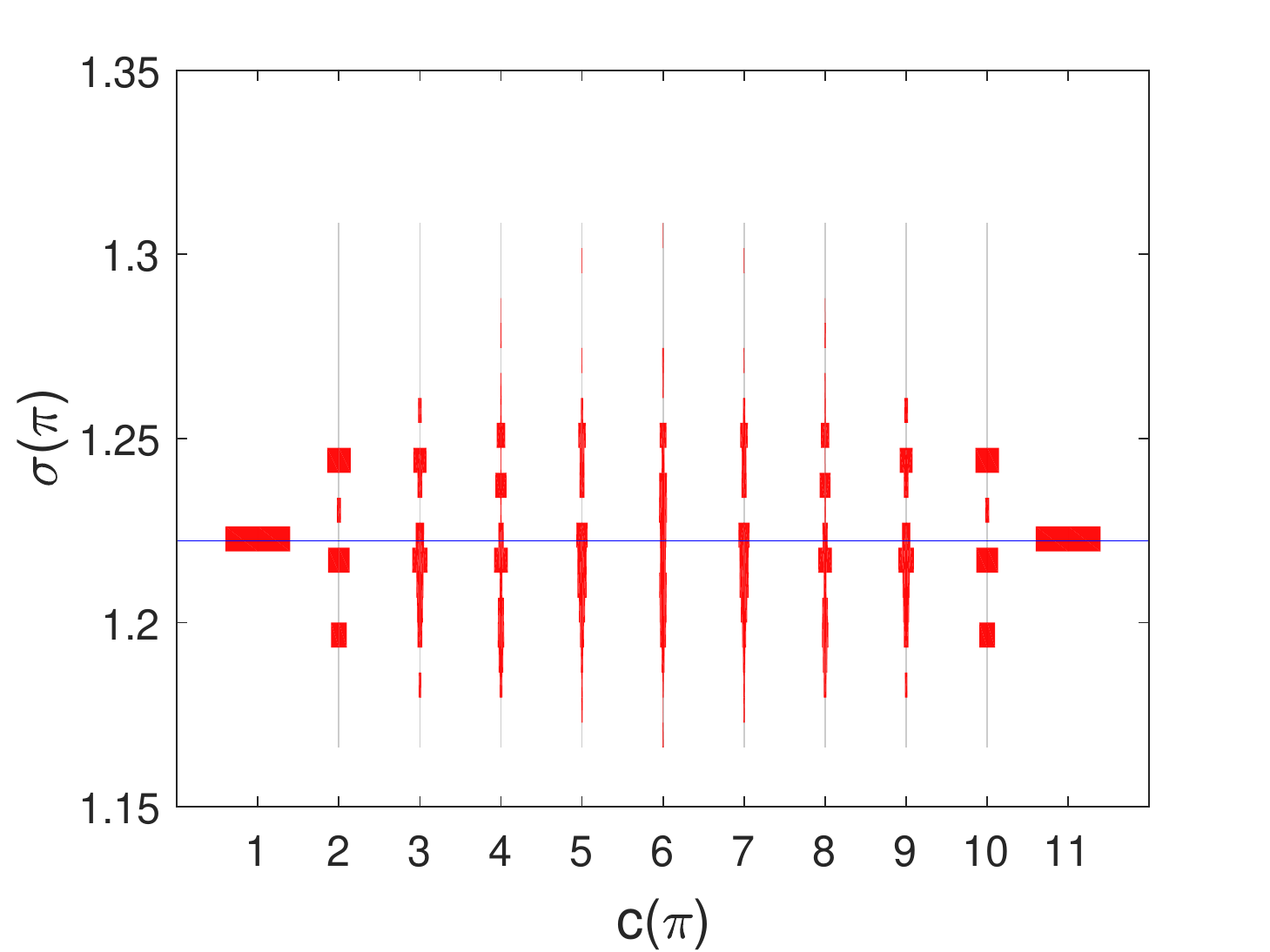}

\hspace{1cm} (a)  \hspace{10cm} (b) 

\caption{Distributions of the structure coefficients $\sigma(\pi)$ for graphs: (a) Frucht, Fig. \ref{fig:graph}(a); (b) Chvatal, Fig. \ref{fig:graph}(e).   The blue lines intersecting the distributions show the values $\sigma=\{3/2,11/9\}$ obtained for a single cooperator with $N=12$, $k=\{3,4\}$ and Eq. (\ref{eq:sig}).} 
\label{fig:viol_frucht_frank}
\end{figure}

We next present and discuss numerical results for the interaction networks given in Fig.  \ref{fig:graph}. As shown above,
condition (\ref{eq:cond1}) depends linearly on the scaling parameters for dilemma strength $u$ and $v$ yielding straight lines with $u=\sigma(\pi)-1-v$ from north--west to south--east in the $uv$--parameter plane, see Fig. \ref{fig:uv_plane}(a).  Such a straight line is the more towards the north--east corner, the larger the structure coefficient $\sigma(\pi)$ is. In other words, the largest $\sigma(\pi)=\sigma_{max}$  gives the largest section in the bounded $uv$--plane generalizing elements of the payoff matrix for which cooperation is favored over defection, while the smallest $\sigma(\pi)=\sigma_{min}$ gives the smallest section. We first briefly look at how the structure coefficients are distributed for each number of cooperators $c(\pi)$, see Fig. \ref{fig:viol_frucht_frank} for the Frucht ($k=3$) and the Chvatal ($k=4$) graph. We notice that for 1 cooperator ($c(\pi)=1$) and 1 defector ($c(\pi)=N-1=11$) we get the single values  $\sigma=\{3/2,11/9\}$ obtained for a single cooperator according to Eq. (\ref{eq:sig}). For $2 \leq c(\pi) \leq 10$ we obtain a symmetric distribution with some values of $\sigma(\pi)$ larger and some smaller than $\sigma$, depending on the arrangement of cooperators and defectors on the evolutionary graph. This means there are for the same number of cooperators $c(\pi)$ some configurations $\pi$ that are more prone to cooperation than others. 

\begin{figure}[htb]

\includegraphics[trim = 0mm 0mm 0mm 0mm,clip,width=8.5cm, height=5.9cm]{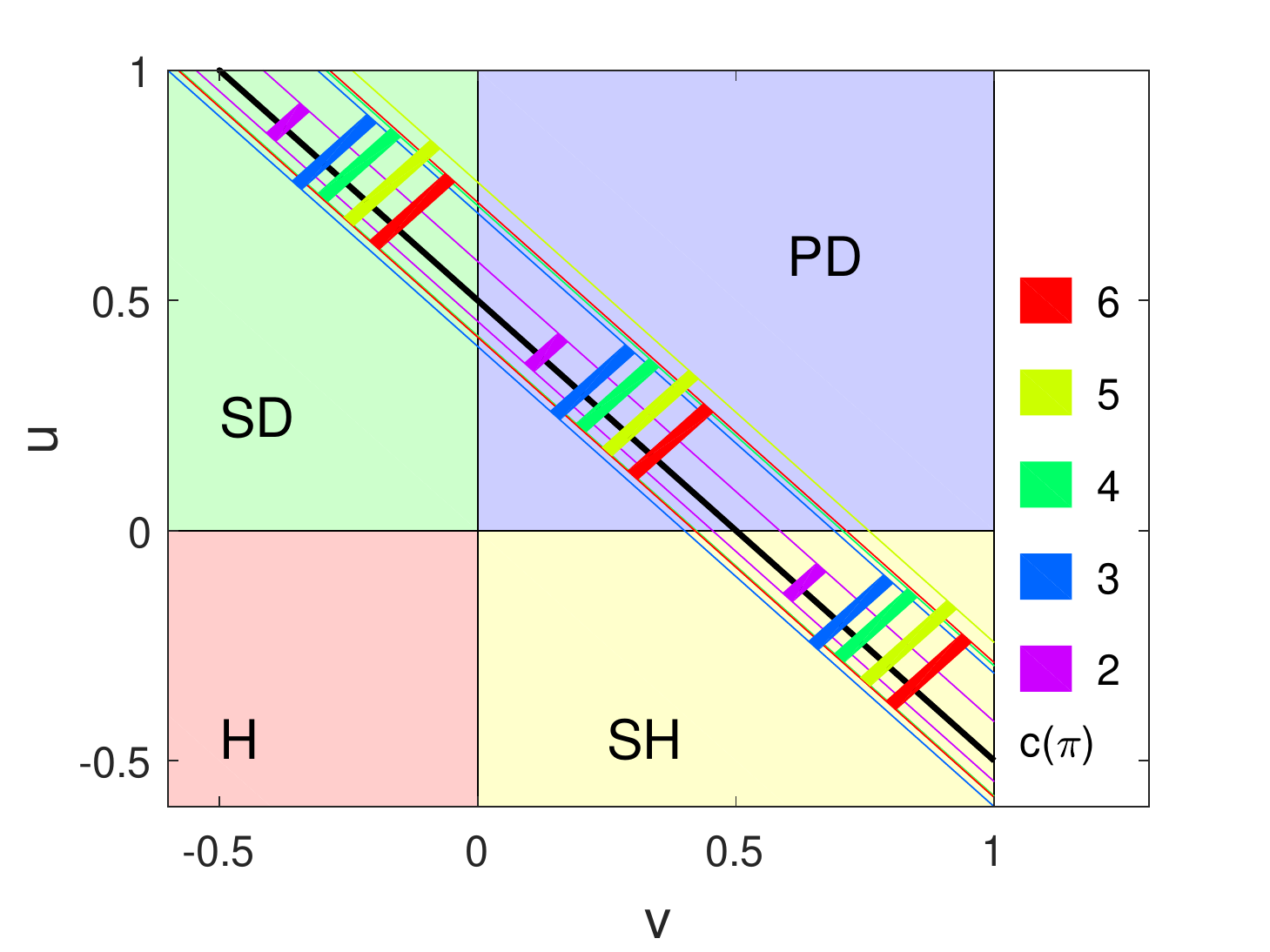}
\includegraphics[trim = 0mm 0mm 0mm 0mm,clip,width=8.5cm, height=5.9cm]{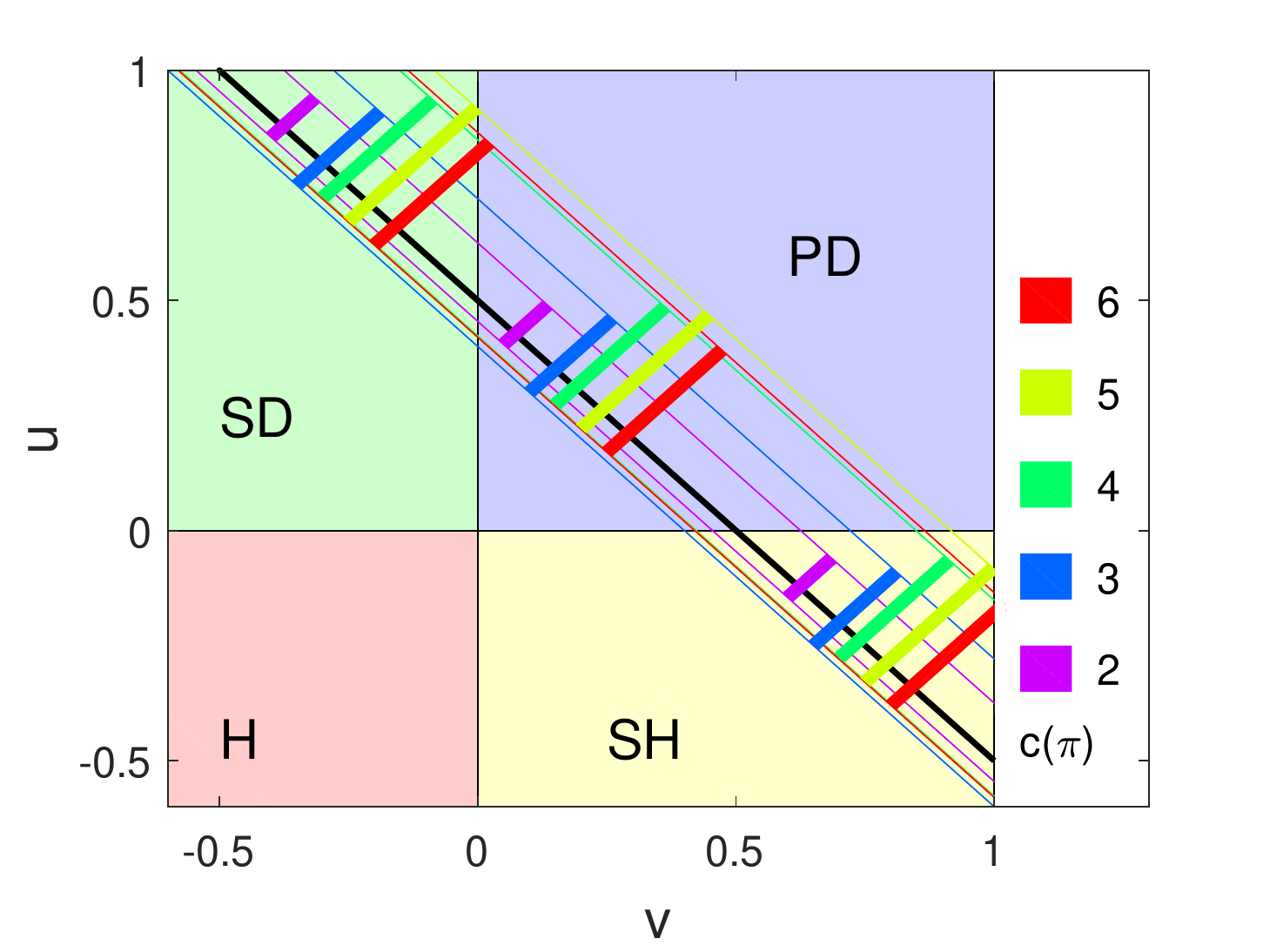}

\hspace{0.5cm} (a) Network in Fig. \ref{fig:graph}(a): Frucht graph \hspace{2cm} (b) Network in Fig. \ref{fig:graph}(c)

\includegraphics[trim = 0mm 0mm 0mm 0mm,clip,width=8.5cm, height=5.9cm]{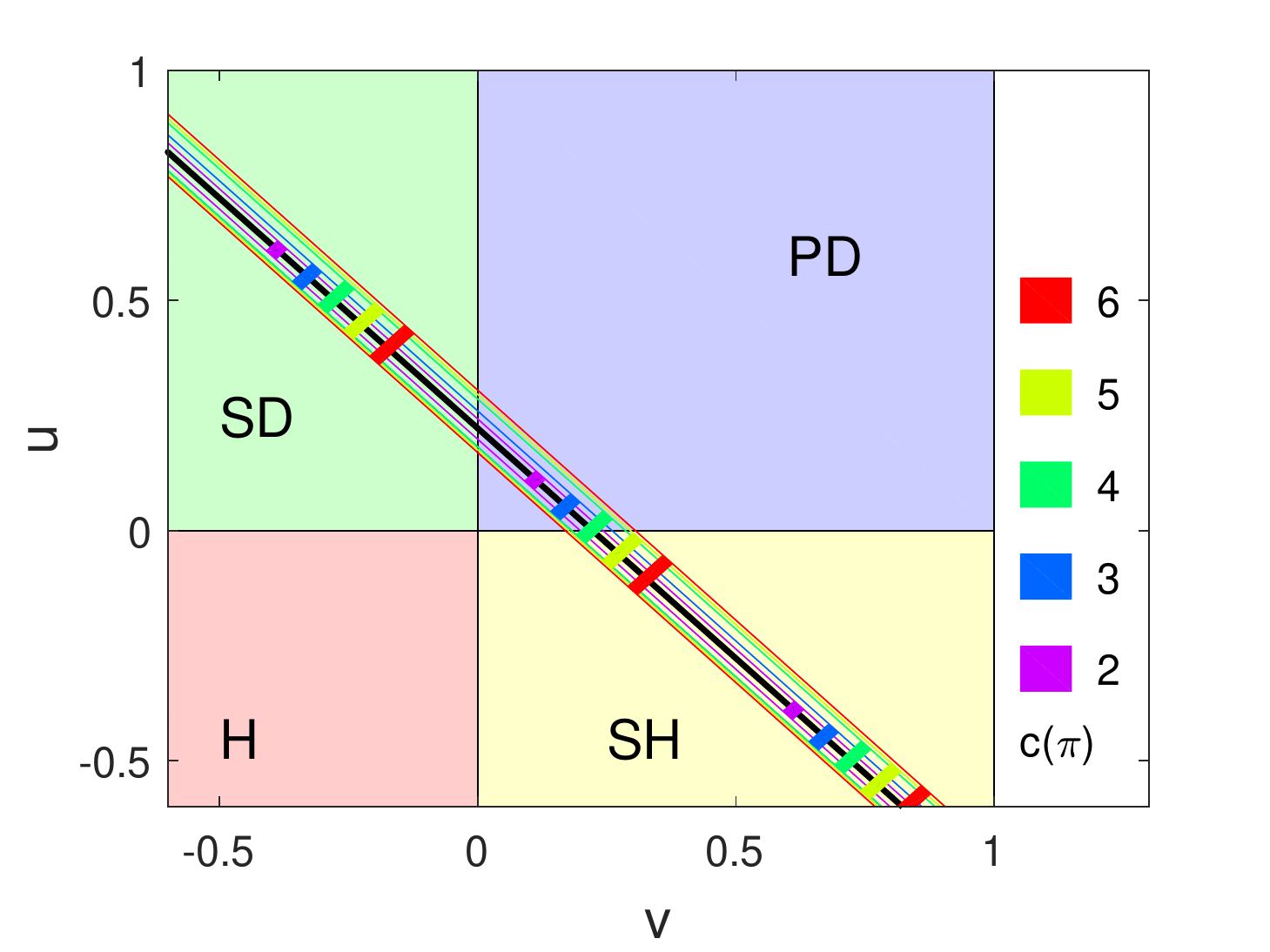}
\includegraphics[trim = 0mm 0mm 0mm 0mm,clip,width=8.5cm, height=5.9cm]{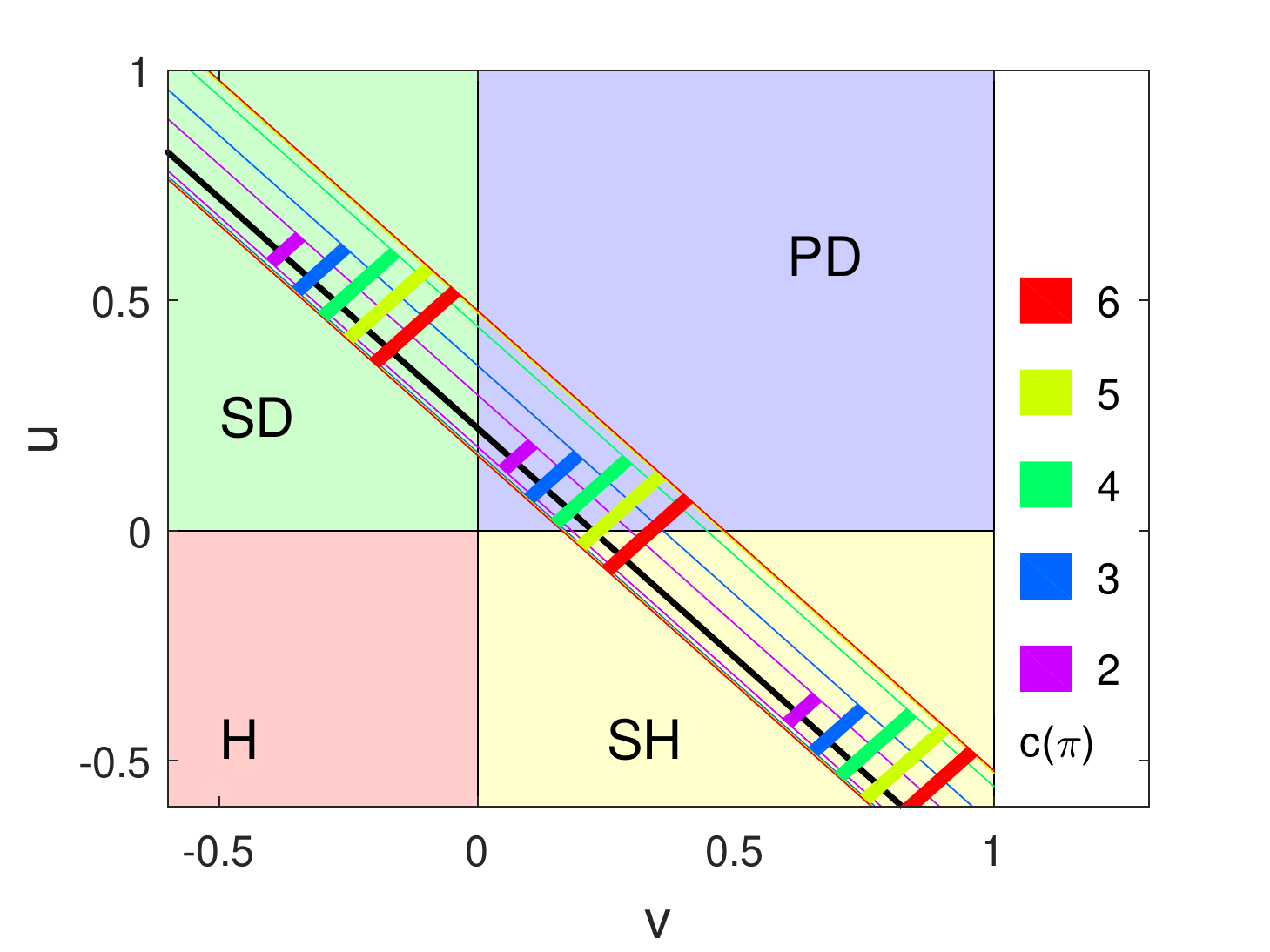}

\hspace{0.5cm} (c) Network in Fig. \ref{fig:graph}(e): Chvatal graph  \hspace{2cm} (d) Network in Fig. \ref{fig:graph}(g)

\caption{Sections of condition $u<\sigma(\pi)-1-v$ between the maximal and minimal structure coefficients, $\sigma(\pi)=\sigma_{max}$ and $\sigma(\pi)=\sigma_{min}$ for the interaction networks given in Fig. \ref{fig:graph} for the number of cooperators $2 \leq c(\pi) \leq 6$. The bars give the range of the condition $u<\sigma(\pi)-1-v$ between $\sigma_{max}$ and $\sigma_{min}$ according to the color code for each number of cooperators $c(\pi)$. The black line indicates the section for a single cooperator with $\sigma$ according to Eq. (\ref{eq:sig}).} 
\label{fig:uv_plane_graph}
\end{figure} 

The main result is given by Fig. \ref{fig:uv_plane_graph}, which shows for the interaction networks given in Fig. \ref{fig:graph} the sections of condition $u<\sigma(\pi)-1-v$ between the maximal and minimal structure coefficients, $\sigma(\pi)=\sigma_{max}$ and $\sigma(\pi)=\sigma_{min}$  for the number of cooperators $2 \leq c(\pi) \leq 6$ (the results for $7 \leq c(\pi) \leq 10$ are omitted as they are symmetric, see Fig. \ref{fig:viol_frucht_frank}).
Different graphs yield different $\sigma_{min}$ and $\sigma_{max}$ for each $c(\pi)$, which in turn produce different ranges in the $uv$--plane indicating that cooperation is favored or not. For instance, the interaction graph given in Fig. 
\ref{fig:graph}(d) has for $c(\pi)=5$ cooperators the largest value $\sigma_{max}=1.9159$. For this configuration $\pi$ cooperation prevails for almost all SD and SH games and a considerable fraction of PD games, see the green--yellow line in Fig. 
\ref{fig:uv_plane_graph}(b). Comparing the interaction graphs reveals that the lower line defined by $\sigma_{min}$ is the same (or almost the same) for the  examples considered, while the upper line is not and gives the largest range for the graph  and configuration in Fig. 
\ref{fig:graph}(d).  However, comparing $k=3$ and $k=4$ shows that the upper and lower lines are more in favor for cooperation for $k=3$ than for $k=4$, which is a consequence of the structure coefficient for a single cooperator$\sigma$ with $\sigma_{min} \leq \sigma \leq \sigma_{max}$ being smaller for $k=4$ as for $k=3$, see Fig. \ref{fig:uv_plane}(b). Compare also to Fig. \ref{fig:viol_frucht_frank} showing that for all configurations the structure coefficients $\sigma(\pi)$  for the Frucht graph ($k=3$) are  larger than for the Chvatal graph ($k=4$). A recent work suggests that it is general result for regular interaction graphs that structure coefficients fall for the degree $k$ ($=$ number of coplayers) of the graph getting larger~\cite{rich18a,rich18b}.
\section{Conclusions}
Different types of social dilemma games such as Prisoners dilemma, stag--hunt or snow--drift can be universally expressed  by a scaling of dilemma strength. 
We have considered structure coefficients defined for each configuration describing any arrangement of cooperators and defectors on a regular evolutionary graphs. As these structure coefficients are linked to whether or not cooperation is favored over defection, we could study how the universal scaling of dilemma strength relates to specific favorable configurations of cooperators and defectors. The main findings are that some graphs  permit certain arrangements of cooperators and defectors to possess particularly large structure coefficients. Moreover, these large coefficients imply particularly large sections of a bounded parameter plane spanned by scaling  gamble--intending and risk--averting dilemma strength. In addition, the sections can be described by linear inequalities depending only on the two scaling parameters.

%
%


\begin{thebibliography}{6}
%

\bibitem{allen14} Allen, B., Nowak, M. A.:  Games on graphs. EMS Surv. Math. Sci. 1, 113--151 (2014)

\bibitem{allen17} 
Allen, B.,  Lippner, G.,  Chen, Y.~T.,   Fotouhi, B.,  Momeni, N.,  Yau, S.~T.,  Nowak, M.~A.:  Evolutionary dynamics on any population structure.
Nature 544, 227--230  (2017)

\bibitem{broom13} Broom, M.,  Rychtar, J.: Game-Theoretical Models in Biology. 
 Chapman and Hall/CRC, Boca Raton, FL  (2013)
 
 \bibitem{chen13}  Chen, Y. T.:  Sharp benefit--to--cost rules for the evolution of cooperation on regular
graphs.
Ann. Appl. Probab. 3, 637--664  (2013)


\bibitem{chen16} Chen,  Y. T.,  McAvoy, A., Nowak, M. A.:  Fixation probabilities for any configuration of two strategies on regular graphs.
Sci. Rep. 6,
39181  (2016)
 
 \bibitem{hinder15}  Hindersin, L.,  Traulsen, A.: Most undirected random graphs are amplifiers of selection for birth--death dynamics, but suppressors of selection for death--birth dynamics. PLoS Comput Biol11, e1004437 (2015)



 
\bibitem{nowak06} Nowak, M. A.:  Evolutionary Dynamics: Exploring the Equations of Life. Harvard University Press, Cambridge, MA (2006)


\bibitem{perc10} Perc, M., Szolnoki, A.:  Coevolutionary games--A mini review.  BioSystems 99, 109--125, (2010)



 





\bibitem{rich16} Richter,  H.:
 Analyzing coevolutionary games with dynamic fitness landscapes. In:  Y. S. Ong  (ed.),   Proc. IEEE Congress on Evolutionary Computation, IEEE CEC 2016,  IEEE Press, Piscataway, NJ,   609--616 (2016)

\bibitem{rich17} Richter, H.: Dynamic landscape models of coevolutionary games.    BioSystems 153--154, 26--44 (2017)


\bibitem{rich18a} Richter, H.: Properties of network structures, structure coefficients, and benefit--to--cost ratios.  BioSystems 180, 88--100 (2019)


\bibitem{rich18b} Richter, H.: Fixation properties of multiple cooperator configurations on regular graphs. Theory of Biosciences 138 (2019), in press.

\bibitem{sha12} Shakarian, P., Roos,  P.,  Johnson, A.: A review of evolutionary graph theory with applications to game theory.   BioSystems 107, 66--80 (2012)




\bibitem{szabo07}   Szabo G., Fath, G.:   Evolutionary games on graphs.  Phys. Rep. 446, 97--216 (2007)

\bibitem{tani09} 
Tanimoto, J.: A simple scaling of the effectiveness of supporting mutual cooperation in donor-recipient 
games by various reciprocity mechanisms.   BioSystems 96, 29--34 (2009) 

\bibitem{tarnita09}
    Tarnita, C. E.,  Ohtsuki, H., Antal, T., Fu, F., Nowak, M.~A.: 
Strategy selection in structured populations. J. Theor. Biol. 259, 570--581 (2009)


\bibitem{taylor07}  Taylor, P. D., Day, T., Wild, G.:     Evolution of cooperation in a finite homogeneous graph. Nature 447, 469--472 (2007)

\bibitem{wang15} Wang, Z., Kokubo, S., Jusup, M., Tanimoto, J.:  Universal scaling for the dilemma strength in evolutionary games.  Phys.  Life Rev. 14, 1--30 (2015)

\bibitem{zuk13} Zukewich, J., Kurella, V., Doebeli, M., Hauert, C.: Consolidating birth--death and death--birth processes in structured populations. PLoS One 8: e54639 (2013) 

\end{thebibliography}
\end{document}